\documentstyle[11pt,twoside,psfig,jltp]{article}

\title{Bcc $^4$He as a Coherent Quantum Solid: "Super-Solid" ?}

\author{Nir Gov\address{Department of Physics, 
University of Illinois at Urbana-Champaign, \\
1110 Green St., Urbana 61801, IL, USA}}

\runninghead{N. Gov}{Bcc $^4$He as a Coherent Quantum Solid: "Super-Solid" ?}
       
\begin{document}

\begin{abstract}
In this work we investigate the quantum nature of bcc $^{4}$He. We show that
it is a solid phase with an Off-Diagonal Long Range Order of coherently
oscillating local electric dipole moments. These dipoles arise from the
correlated zero-point motion of the atoms in the crystal potential, which
oscillate in synchrony so that the dipolar interaction energy is minimized.
This coherent state has 
a three-component complex order parameter. The
condensation energy of these dipoles in the bcc phase further stabilizes it
over the hcp phase at finite temperatures. This condensation of the dipoles is not a
'super-solid'. We further
show that there can be fermionic excitations of this ground-state and
predict that they form an optic-like branch in the (110) direction.

\end{abstract}

\maketitle


\section{INTRODUCTION}

The bcc phase of $^{4}$He has a pronounced quantum nature due to
the relatively open structure of the lattice. Quantum effects are
manifested in strong anharmonicity of some phonon modes and in the
large zero-point kinetic energy of the atoms \cite{glyde}.
In a previous paper \cite{niremil}
we have proposed a new physical model for the local atomic zero-point motion
in the bcc phase. In this model we assume that
there exist in the bcc
$^{4}$He a phase with coherently oscillating and anisotropic local
electric dipoles. The ground-state of these coherent dipoles
minimizes the dipolar interaction energy between them. 
We find that
bosonic phase fluctuations in the (110) direction reproduce the
spectrum of the anomalously soft T$_{1}(110)$ phonon.
Local dipolar flipping results in a new optic-like branch, described using Fermi statistics.

\section{GROUND-STATE COHERENT DIPOLES}

In our model \cite{niremil} we focused on the effects of the local
zero-point motion of the atoms inside the potential-well on the nature of the
ground-state.
Since in the directions
normal to the unit cube's faces (i.e. (100),(010) etc.) the
confining potential well of an atom due to its neighboring atoms
is very wide with a pronounced double-minimum structure (Fig.1),
the dynamic correlations between the atoms along these directions will tend to keep them apart \cite{nosanow}, by
virtually exciting the atom to oscillate between the minima of the
potential along the (100) direction, correlated with similar
zero-point motions of the other atoms (quantum resonance).
\begin{figure}
\centerline{\psfig{file=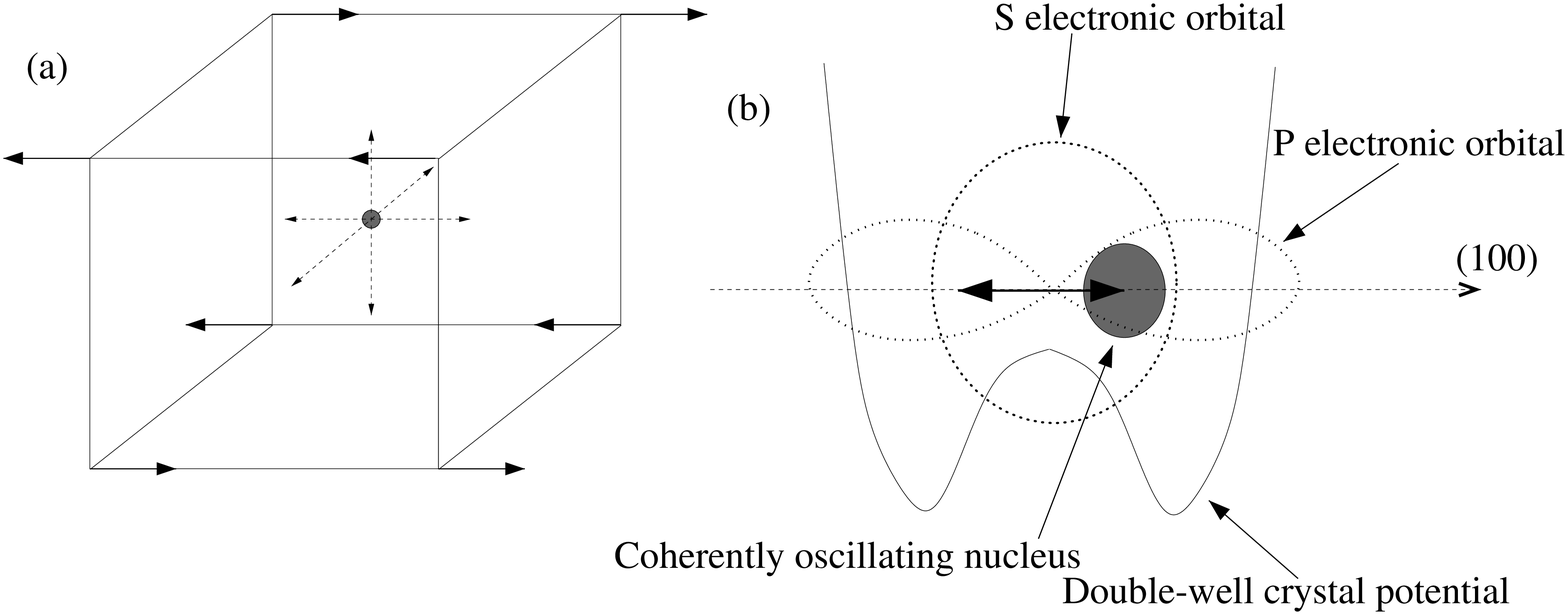,height=2.4in,width=5.2in}}
\caption{(a) The coherent zero-point motion executed by the atoms along the normal directions in the crystal potential. The large arrows show the configuration of the oscillating dipoles along one of the normal axes. (b) The coherent oscillations of the ion in the double-well potential distorts the electronic cloud. This creates the oscillating electric dipole moments.}
\end{figure}
If we relax the Born-Oppenheimer approximation, and allow some
relative motion between the ions and the electrons, we will obtain
that the motion of the ions creates an oscillating electric dipole.
This oscillating electric
dipole moment ${\bf \mu }$, is equivalent to a mixing of the $s$
and $p$ electronic levels \cite{niremil} (Fig.1). Since these zero-point dipole moments are correlated between the different atoms, the resulting dipolar interactions are minimized by the configuration of Fig.1a.
It is the dipolar interactions that drive the correlated zero-point nuclear oscillation with energy $E_{0}$.
The oscillating state of quantum resonance of each atom is
of the form: 
$\left| \Psi (t)\right\rangle =\left| s\right\rangle +\lambda
e^{iE_{0}t/\hbar }\left| p_{x,y,z}\right\rangle$,
 where $\left| s\right\rangle $,$\left| p_{x,y,z}\right\rangle $ are the
electronic ground-state and first excited state, and $\lambda \simeq 0.01$ being the
mixing of the levels due to the oscillation of the nucleus.
The system
oscillates in resonance between the equivalent up/down arrangements of the
ground-state dipoles (Fig.1), 
simultaneously arranged along all three
orthogonal axes.

The effective
Hamiltonian for the interacting coherent dipoles \cite{niremil,hopfield,anderson}

\begin{equation}
{H_{loc}} ={\sum_{k}}(E_{0}+X(k))\left( {{b_{k}}^{\dagger }}{b_{k}}\right)
+{\sum_{k}}X(k)\left( {{b_{k}}^{\dagger }}{b_{-k}^{\dagger }}%
+b_{k}b_{-k}\right)  \label{hloc}
\end{equation}
where ${{b_{k}}^{\dagger },}{b_{k}}$ are Bose creation/annihilation operators
of the local mode, $E_{0}$ is the energy of exciting a local dipole out of
the correlated ground-state, and $X(k)$ is \cite{heller}:
$X\left( {\bf k}\right) =-\left| {\bf \mu }\right| ^{2}\sum_{i\neq 0}\left[
\frac{3\cos ^{2}\left( {\bf \mu }\cdot \left( {\bf r}_{0}-{\bf r}_{i}\right)
\right) -1}{\left| {\bf r}_{0}-{\bf r}_{i}\right| ^{3}}\right]  
\times \exp \left[ 2\pi i{\bf k}\cdot \left( {\bf r}_{0}-{\bf r}_{i}\right)
\right]$, 
where ${\bf \mu }$ is the oscillating electric dipole moment,
perpendicular to the wavevector ${\bf k}$ which is 
along the (110) (or equivalent) direction, where modulations of
the array of dipoles shown in Fig.1 correspond to phonons of the
lattice \cite{niremil}.

The Hamiltonian ${H_{loc}}$ (\ref{hloc}) which describes the effective
interaction between localized modes can be diagonalized using the usual Bogoliubov
transformation.
The new ground-state 
is a coherent state,
where excitations have the energy spectrum:
$E(k)=\sqrt{E_{0}\left( E_{0}+2X(k)\right) }$.

The bare energy $E_{0}$ to flip a local-dipole out of the coherent
ground-state is $E_{0}=-2X(0)$, i.e. twice the ground-state dipolar interaction energy. The empirical value of $E_{0}\simeq 7$K
taken from NMR data \cite{schuster} therefore fixes the size of the coherent dipole moment $%
\mu $.
Since the phonon is a modulation of the relative phases of the atomic motion, it is therefore equivalent to the excitation described by $E(k)$.
The agreement we find by comparing $E(k)$ with the transverse T$_{1}$(110) phonon data
taken from inelastic neutron scattering \cite{niremil}, emphasizes the
self-consistency of our description. 
Further comparisons of this model
with experimental data can be found in \cite{niremil}.

\section{ODLRO AND CONDENSATION}

The coherent ground-state defines a global phase and
breaks the global gauge symmetry of a well-defined occupation number of local
dipoles. In the limit $k\rightarrow 0$ we find that the occupation number of
the local-modes diverges as $1/k$, signaling macroscopic Bose-Einstein
condensation in the zero-momentum state: 
$\left\langle n_{k}\right\rangle \rightarrow _{k\rightarrow 0}\frac{1%
}{2}\frac{E_{0}/2}{E(k)}=\frac{E_{0}/2}{2\hbar kc}$,
where $c$ is the sound velocity of the T$_{1}$(110) phonon which is the
natural excitation of the dipolar array. This is just the order of
divergency in the occupation number which is typical of an interacting Bose
system \cite{gavoret}, where it can be related to the occupation of the
zero-momentum state, i.e., the condensate fraction $n_{0}/n$:
 $\left\langle n_{k\rightarrow 0}\right\rangle =\frac{n_{0}}{n}\frac{mc}{%
2\hbar k}$,
where $m$ is the boson mass.
Since we describe local excitations whose total number is not a conserved
quantity, their condensate fraction can not be simply defined. Nevertheless
the coherent dipoles do correspond to the behavior of the $^{4}$He atoms so comparing the residues of the divergent part in these two expressions we find 
an effective condensate fraction:
 $\frac{n_{0}}{n}=\frac{E_{0}/2}{mc^{2}}\simeq \frac{3.5}{10}=35\pm 8\%$, 
where $c$ is \cite{minki}
$\sim 130-160$m/sec.
We note that the
condensation of the local modes is only in the directions (110).
This means that the
limited (vanishing in the T=0 case) phase space volume will decrease the
overall condensate fraction. At the bcc temperatures ($\sim $%
1.4K) the one dimensional chains in the (110) directions are thermally
broadened so that the condensation is now over finite volume sections of
phase space. We can estimate this effect,
and find a condensate fraction:
$n_{0}/n\sim 0.5\%$. 

Similar condensation of local dipoles along all three orthogonal axes of local
zero-point motion means that there are three independent phases at each lattice site.
The order-parameter 
can be described as a vector of three complex functions of
independent phase: 
 $\Phi ({\bf r})=\left(
\left| \mu \right| e^{i\theta _{x}({\bf r})} ,
\left| \mu \right| e^{i\theta _{y}({\bf r})} ,
\left| \mu \right| e^{i\theta _{z}({\bf r})}
\right)$, where $\left| \mu \right|$ is the size of the coherent dipole moment. The bcc $^{4}$He is therefore a system having both Diagonal Long Range Order
(DLRO) of the solid lattice and Off-Diagonal Long Range Order (ODLRO) of the
local dipoles. It is not a 'super-solid' \cite{andreev}
in that it does not contain both a superfluid and a solid, but is more
similar to the superconductors which have a DLRO of the atoms in the lattice
and ODLRO\ of the superconducting electrons \cite{kohn}.
Recent experiments
\cite{kanorsky} on the behavior of implanted metallic atoms (Cs) in solid $%
^{4}$He reveal strong coherence effects in the bcc phase,
in accordance with our expectation of a coherent state of dipoles.

The condensation energy of the dipoles
lowers the energy of the
ground-state of the bcc phase and further stabilizes it with respect to the
hcp phase \cite{anderson}:
$\Delta E=\sum_{k}\frac{E(k)-(E_{0}+X(k))}{2}<0$
, which is negative since $X(k)<0$.
We can
estimate this sum at the finite temperature of the bcc
phase: 
$\Delta E\simeq -2$mK per atom. This result is
in agreement with the experimentally interpolated energy difference between
the bcc and hcp phases \cite{balibar} of solid $^{4}$He, which is of the
order of a few mK per atom.

\section{FERMIONIC EXCITATIONS}

In addition to the fluctuations of the phase of the coherent dipole
ground-state (i.e. T$_{1}$(110) phonons), there can be a 'dipole-flip' mode,
 where a dipole is in anti-phase (phase difference of $\pi $) relative
to the ground-state configuration of the dipoles. As we
mentioned before, this is just the definition of the bare local-mode
energy $E_{0}$. Such an excitation is naturally treated as a fermion since a
flipped dipole is antisymmetric with respect to its ground-state
configuration, that is with respect to the global phase
of $\Phi ({\bf r})$.
The effective Hamiltonian describing such a fermion should contain a
term describing the creation and annihilation of pairs of fermions from the
ground-state by a T$_{1}$(110) phonon. In
addition there is a term that describes the excitation energy of the
bare fermionic localized-mode, i.e. $E_{0}$.
The effective Hamiltonian that we therefore propose is
\begin{equation}
H_{D}=\sum_{k}E(k)\left( c_{k}^{\dagger }c_{-k}^{\dagger
}+c_{k}c_{-k}\right) -E_{0}\sum_{k}c_{k}^{\dagger}c_{k}  \label{diracham}
\end{equation}
where $c_{k}^{\dagger },c_{k}$ are the creation/annihilation operators
of the flipped dipoles.
In the absence of the second term we have just
the Bose ground-state rewritten in terms of fermion pairs.
We linearize and solve the equations of motion that follow from (\ref{diracham}),
similar to the BCS method \cite{kittel},
with the resulting energy spectrum:
 $E_{f}(k)=\sqrt{E_{0}^{2}+E(k)^{2}}$.
\begin{figure}
\centerline{\psfig{file=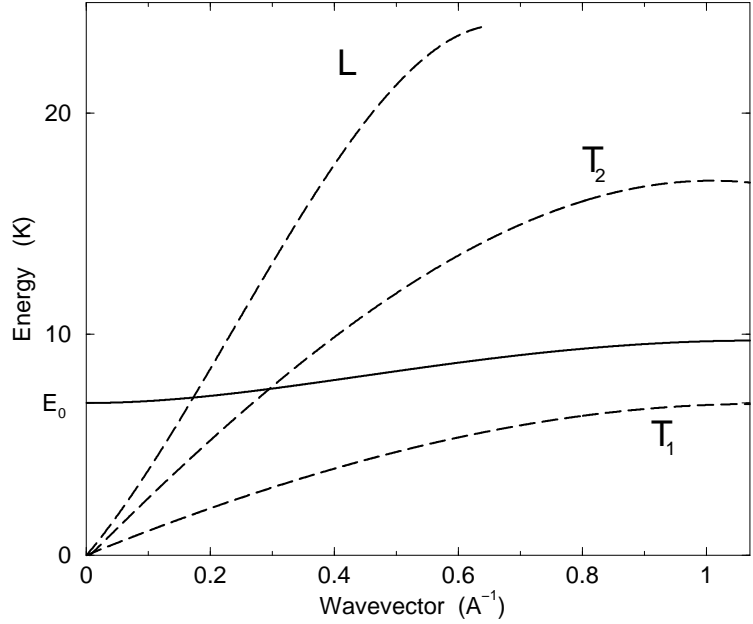,height=3.3in}}
\caption{The spectrum of the fermionic optic-like mode ($E_{f}(k)$, solid
line) compared with the experimentally measured phonons in the (110)
direction [9].}
\end{figure}
In Fig.2 we plot the energy spectrum $E_{f}(k)$ compared with the other
phonon modes in the (110) direction \cite{minki}. 
No other optic-like phonons are
expected in the bcc lattice.
These predictions await high-resolution neutron, Raman and x-ray
scattering experiments to be compared with.
Note also that these excitations are confined to the (110)
directions so despite their relatively low energy, they do not
contribute significantly to the specific heat of the solid.

\section{CONCLUSION}

We have identified a three component complex order parameter
and Bose-Einstein condensation in the bcc solid phase, though not a
'super-solid' \cite{kohn}. There can be
further manifestations of the ODLRO of the dipoles in the bcc phase which we
have not explored yet, such as macroscopic topological defects in this 
complex order-parameter. 
We further obtain that a locally flipped-dipole 
will behave as a fermion, 
with an optic-like branch in the (110) direction.

\section*{ACKNOWLEDGMENTS}
I thank Emil Polturak for useful discussions and encouragement.
This work was supported by 
the Fulbright Foreign Scholarship grant, 
the Center for Advanced Studies and
NSF grant no. PHY-98-00978.

\end{document}